\documentclass[useAMS,usenatbib]{mn2e}
\usepackage{epsfig,psfig}
\title[Millimetric properties of GRB host galaxies]
{Millimetric properties of gamma ray burst host galaxies}
\author[Priddey et al.]{R.S. Priddey$^{1}$
\thanks{E-mail: priddey@star.herts.ac.uk},
N.R. Tanvir$^{1}$, A.J. Levan$^{1}$, 
A.S. Fruchter$^{2}$,
C. Kouveliotou$^{3}$, \and
I.A. Smith$^{4}$, 
R.A.M.J. Wijers$^{5}$
\\
$^{1}${\it Centre for Astrophysics Research, University of Hertfordshire,
College Lane, Hatfield AL10 9AB, UK}\\
$^{2}${\it Space Telescope Science Institute, 3700 San Martin 7 Drive,
Baltimore, MD21218, USA}\\
$^{3}${\it NASA Marshall Space Flight Center, NSSTC, SD-50, 
320 Sparkman Drive, Huntsville, 
AL 35805, USA}\\
$^{4}${\it Dept. of Physics \& Astronomy, Rice University MS108, Houston,
TX 77005-1892, USA}\\
$^{5}${\it Astronomical Institute 'Anton Pannekoek', 
Faculty of Science, University of Amsterdam, Kruislaan 403,}\\
{\it 1098 SJ Amsterdam, The Netherlands}
}
\begin{document}

\date{MNRAS, in press}

\pagerange{\pageref{firstpage}--\pageref{lastpage}} \pubyear{2006}

\maketitle

\label{firstpage}

\begin{abstract}

\noindent
We present millimetre (mm) and submillimetre (submm) photometry
of a sample of five host galaxies of Gamma Ray Bursts (GRBs),
obtained using the MAMBO2 and SCUBA bolometer arrays respectively. 
These observations were obtained as part of an ongoing project to
investigate the status of GRBs as indicators of star formation.
Our targets include two of the most unusual GRB host galaxies, selected
as likely candidate submm galaxies: the extremely red 
($R-K\approx5$) host of GRB 030115, and the extremely faint ($R>29.5$) 
host of GRB 020124. Neither of these galaxies is detected, but the deep
upper limits for GRB 030115 impose constraints on its spectral
energy distribution, requiring a warmer dust temperature than is 
commonly adopted for submillimetre galaxies.

As a framework for interpreting these data, and for predicting the
results of forthcoming submm surveys of {\it Swift}-derived host samples,
we model the expected flux and redshift distributions based on
luminosity functions of both submm galaxies and GRBs, assuming a
direct proportionality between the GRB rate density and the global
star formation rate density. We derive the effects of possible sources
of uncertainty in these assumptions, including (1) introducing
an anticorrelation between GRB rate and the global average metallicity, 
and (2) varying the dust temperature.

\end{abstract}


\section{Introduction}
There is now strong evidence linking long-duration 
Gamma Ray Bursts (GRBs) with the core-collapse of massive stars
(e.g. Hjorth et al. 2003a)--- and hence, given the short main-sequence
lifetimes of such stars, with star formation activity.
Indeed, as tracers of star formation, GRBs hold a number of advantages 
over traditional methods. The high luminosity  of their prompt emission
and afterglows enable them to be detected, in principle, out to redshifts
$\ga$10 (in practice currently out to $z>6$, e.g. Haislip et al. 2006).
Their high energy emission can furthermore pass unaffected
through intervening gas and dust--- the very conditions one would expect
to be associated with massive star formation.
As the outcome of a single stellar event,
the luminosity of the GRB ought to be independent of that
of its host galaxy, 
enabling localisation of galaxies too faint, dusty or
distant to be detected by traditional means, thus sidestepping many of the
biases that afflict optical and submm surveys. 
Furthermore, 
spectroscopy of the bright optical afterglows enables 
one to measure the redshift and other properties of the host galaxy, 
even when direct detection of the galaxy may be infeasible 
(e.g. Berger et al. 2002).

Once the star-forming properties of a carefully-selected subsample
of GRB hosts have been established, it should be possible to derive the
the star formation history of the Universe, by measuring the redshift
distribution of GRBs.
A purely GRB-selected galaxy sample should, furthermore, represent
an unbiased census of the galaxy types reflecting their
relative contribution to the bulk star formation rate. 
Hitherto it has been difficult to assess the biases afflicting 
the assembly of such samples, 
factors modulating the GRB rate
as a function of redshift (for example, 
redshift dependence of density of surrounding medium; metallicity;
stellar initial mass function (IMF);
the distribution of jet opening angles).
Follow-up of samples of bursts detected with
{\it Swift} (Gehrels et al. 2004) 
shows promise in being able to characterise and
overcome such biases. 
The BAT (Burst Alert Telescope) 
detector is more sensitive to high-redshift bursts
than previous missions (e.g. Band 2006), and,
due to the rapid localisation of bursts via the onboard XRT (X-Ray Telescope),
ground-based follow up of afterglows (yielding information constraining
the physical properties of the afterglow, for example 
redshift, spectral slopes, light curves and 
jet-break times) is much more systematic.  

However, the true proportionality between the global GRB and star formation
rates has yet to be definitively established. It is plausible, for instance,
that an otherwise direct relation is complicated by dependence on conditions
local to the GRB. For example, it is thought that metallicity could
play a role in the GRB formation process (e.g. Fynbo et al. 2003; Fruchter
et al. 2006).
Before one can begin to exploit the new GRB catalogue produced by 
a mission such as {\it Swift}, 
it is of the utmost importance to characterise such effects.

One of the most significant contributors to the star formation rate density
at high redshift is the submillimetre galaxy (SMG) population
(e.g. Smail, Ivison \& Blain 1997; Hughes et al. 1998; 
Scott et al. 2002; Borys et al. 2003; Mortier et al. 2005).
Surveys with submm/mm bolometer arrays such as 
SCUBA (Submillimetre Common-User Bolometer Array)
on the James Clerk Maxwell Telescope (JCMT), and
MAMBO (Max Planck Millimetre Bolometer) on the IRAM 
(Institut de Radioastronomie Millim\'{e}trique)
30m telescope,
have revealed a population of galaxies forming the bulk of their
stars in regions optically thick with dust, such that most of their
extremely high luminosity ($L\sim10^{12-13}$L$_{\odot}$) 
is emitted in the rest-frame far infrared region.
Fits to the submm source counts and the FIR background imply that
the integrated power reprocessed by dust contributes a major
fraction of the total luminous energy emitted throughout cosmic history
(Blain et al. 1999).

Taken together, these facts suggest, all other things being 
equal\footnote{One way in which they may {\it not} be equal 
would be, for example,
if a higher than anticipated fraction of SCUBA galaxies were powered by AGN}, 
that highly star-forming 
SMGs ought to yield a high rate of GRBs; conversely, that a high 
fraction of GRB host galaxies should turn out to be luminous 
submillimetre sources. Follow up of GRB hosts in the millimetre
and submillimetre bands therefore
provides one of the most important calibrators of the role of GRBs as 
star formation indicators.

For this reason, in recent years, there have been a number of targetted 
submm studies of GRB hosts, predominantly carried out at 850$\mu$m using
SCUBA on the JCMT (Berger et al. 2003, Smith et al. 1999, 2001, 
Barnard et al. 2003). 
An overview and an analysis is provided by Tanvir et al. (2004). They 
compared observations made with JCMT/SCUBA with model predictions made 
assuming a direct proportionality between the GRB rate and the star 
formation rate. They discovered that, relative to these predictions,
observations show a deficit of bright ($\ga$4mJy) sources. 
Although statistically
only marginally significant, this effect could, if confirmed, have important
ramifications for derivations of the global star formation history based on
GRB surveys.
In all, only three GRB hosts (GRB 000210, $z$=0.85; GRB 000418, $z$=1.12;
GRB 010222, $z$=1.48) out of 23 with 850$\mu$m RMS values $<$1.4mJy 
(Tanvir et al. 2004)
have been securely detected in the submm.
They all lie around the 3mJy level at 850$\mu$m, 
which, though faint, nevertheless implies dust-rich 
($M_{\rm d}\sim10^{8-9}$M$_{\odot}$), massively star-forming
(SFR$\sim$100--1000M$_{\odot}$yr$^{-1}$).
On the other hand, optical/NIR photometry of these hosts reveals
stellar populations that suffer little dust extinction and have low
star formation rates ($\sim$1--10M$_{\odot}$yr$^{-1}$) 
(Gorosabel et al. 2003).
This illustrates the importance of submm observations in understanding
the properties of GRB hosts: relying on optical data alone, their
true significance could easily be overlooked. 
It is thus important to 
study wider samples of hosts, and especially to hunt for additional detections,
to test whether the existing submm-bright sample is typical or unusual.
We are in the process of obtaining X-ray observations of the submm-detected 
sample which should settle this question.)

The present paper pursues these ideas further.
The paper falls into two parts.
First, we present new mm and submm observations. The former were
obtained using the MAMBO2 bolometer array on the 30m IRAM Pico Veleta
telescope, as the preliminary part of the first millimetric survey
explicitly targetting GRB hosts\footnote{Note that the host of GRB 010222
was serendipitously detected at 1.2mm during a mm search for its 
{\it afterglow}}.
The submm data (850 and 450$\mu$m ) were obtained using SCUBA on the JCMT. 
The targets include two of the most extreme GRB hosts known (the reddest and
the faintest), deliberately selected
as the most promising candidate submm galaxies.
Of particular interest, we have obtained deep photometry,
at all three wavelengths (450/850/1200$\mu$m), 
of the reddest afterglow/host found to date, GRB 030115.

In the second part of this paper, we develop models to constrain the 
relation between GRBs and their host galaxies. Using models of the 
luminosity distribution
and evolution of submm galaxies, we derive fits to the luminosity function
of GRBs under the assumption that the GRB rate is a function of the
star formation rate/far-infrared luminosity of the galaxy. 
Based upon these fits, we estimate the flux distributions 
expected at mm and submm wavelengths, which will facilitate comparison
between models of the cosmic star formation history, with future 
mm and submm surveys of the host galaxies of GRBs detected
by {\it Swift}.

\section{Observations}

\subsection{SCUBA/JCMT observations}
The question arises, from previous work, 
as to whether any GRB host galaxies are similar to SMGs
For example, the three submm-detected hosts all have bluer colours than
typical of submm galaxies.
To address this question,
we first used JCMT/SCUBA to target a specific
host whose properties indicate it to be a promising candidate dust-rich,
submm galaxy.
GRB 030115 has the reddest optical colours measured for a GRB host,
implying the presence of a large mass of dust, and a 
(photometric) redshift ($z$=2.5) placing it near the peak of
the redshift distribution measured for submm galaxies (Chapman et al. 2003). 
In these respects, this object constrasts markedly with the three 
submm-detected hosts, all of which have $R-K<3$ and $z<1.5$.

We obtained new 850$\mu$m and 450$\mu$m observations of GRB 030115 with 
JCMT/SCUBA
on 2005 January 27 and 28. SCUBA was used in standard photometry
mode, with a chop throw of 60 arcsec in azimuth. 
The zenith opacity was measured
via skydips and the JCMT water vapour monitor, and 
remained within the range $0.065<\tau_{\rm 225 GHz}<0.08$ on 20050127
and $0.055<\tau_{\rm 225 GHz}<0.06$ on 20050128.
Flux calibration was obtained from the planets Uranus and Mars and several
secondary calibrators.
Data were reduced both manually using the {\sc SURF} package, 
and via the {\sc ORAC-DR} pipeline. Additional sky removal was achieved
by using off-source bolometers to estimate the background.

GRB 030115 had previously been observed by JCMT/SCUBA 
in Target of Opportunity mode commencing 2003 January 18, 
3.3 days after the burst, for a total of two hours:
an upper limit of 6mJy (3$\sigma$), at 850$\mu$m, was reported by
Hoge et al. (2003). Since no afterglow was detected,
we can use this measurement
as an additional upper limit on the submm flux of the host galaxy.
We re-reduced the archived data in the same manner as
described above, to find fluxes as reported in Table 1.

\begin{table*}
\caption{Summary of JCMT/SCUBA observations of GRB 030115. Zenith opacities
are shown at 225GHz, as a range appropriate to the time of observations.}
\begin{tabular}{lcccc}\\
UT date & Observation time & $\tau_{\rm 225 GHz}$ & 
850$\mu$m flux & 450$\mu$m flux \\
& (min) & & \multicolumn{2}{c}{(mJy)} \\
\hline
20050127 & 40 & 0.068--0.070 & 3.2$\pm$1.9 & 71$\pm$34 \\
20050128 & 120 & 0.055--0.058 & $-$1.0$\pm$1.1 & $-$1$\pm$12 \\

20030118 & 115 & 0.083--0.086 & 0.0$\pm$1.7 & 9$\pm$65\\

\end{tabular}
\end{table*}

\subsection{IRAM-30m/MAMBO2 1.2mm data}
Using the 117-element Max Planck Millimetre
Bolometer (MAMBO) detector on the Institut de
Radiostronomie Millim\'{e}trique (IRAM) 30m Pico Veleta 
telescope, we obtained observations of five GRB hosts between December 2004
and April 2005, via pooled (service) observing mode.
Selection of the targets was designed to improve the redshift
distribution of the overall submm/mm GRB host sample, in particular to
try to eliminate a possible bias toward low redshift (see Section 3.2.2).
All the observed targets lie at $z>1$, and their mean redshift is 2.1.

Observations were carried out at a wavelength of 1.2mm
using MAMBO2 in On--Off mode.
Sky opacity was monitored frequently by performing skydips;
regular pointing and focus checks were carried out; and
flux calibration was obtained from standard sources.
The data were reduced using the 
{\sc NIC} software package, which forms part of the
{\sc GILDAS} distribution\footnote{http://www.iram.fr/IRAMFR/GILDAS/}. 
The principles are similar to the SCUBA data reduction described above,
for example the use of off-source bolometers to facilitate sky subtraction.
Details of the observations, and final fluxes of the GRB hosts, are reported in
Table 2.

\begin{table*}
\caption{Details of our new MAMBO2 1.2mm observations of GRB hosts.
Zenith opacities are shown as a range. Quoted fluxes are the 
final values obtained by coadding all the datasets}
\begin{tabular}{lccccc}\\
GRB & $z$ & UT Date & Observing Time & 
Opacity ($\tau$) & Final 250GHz flux density\\
    &     & & (min) & & (mJy)\\
\hline
GRB 020124 & 3.20 & 20050201 & 60 & 0.17--0.19 & 0.28$\pm$0.60\\
           &  & 20050224 & 30 & 0.17--0.18 & \\
           &  & 20050226 & 30 & 0.25--0.34 & \\
           &  & 20050327 & 30 & 0.30--0.35 & \\
\\
GRB 021211 & 1.01 & 20050223 & 30 & 0.09--0.10 & 0.07$\pm$0.53\\
           &  & 20050224 & 30 & 0.18--0.20 & \\
           &  & 20050226 & 30 & 0.25--0.34 & \\
           &  & 20050327 & 30 & 0.30--0.35 & \\
\\
GRB 030115 & 2.5 & 20050223 & 25 & 0.14--0.15 & 0.01$\pm$0.76\\
           &  & 20050224 & 30 & 0.17--0.18 & \\
\\
GRB 030226 & 1.98 & 20050120 & 60 & 0.26--0.28 & $-$0.29$\pm$0.66\\
           &  & 20050224 & 30 & 0.10--0.12 & \\
\\
GRB 030227 & 1.6 & 20041217 & 95 & 0.17--0.20 & $-$0.54$\pm$0.53\\
           &  & 20050226 & 30 & 0.25--0.34 & \\
           &  & 20050327 & 30 & 0.24--0.30 & \\

\hline

\end{tabular}

\end{table*}

\subsection{Results}

None of the hosts is detected, either at 1.2mm with MAMBO or at 850/450$\mu$m
with SCUBA. Moreover, the stacked,
inverse-variance--weighted mean flux of our sample of five new 1.2mm 
observations 
is $-$0.11$\pm$0.27, consistent with a zero flux for this sample. 
For comparison, the weighted mean 850$\mu$m flux of the
sample discussed in Tanvir et al. (2004) is 0.93$\pm$0.18,
which could
be interpreted as a true measure of the flux of the ``typical'' GRB host
(though as discussed by Tanvir et al., the weighted mean carries an 
``observer bias'' in that sources with higher fluxes tend to be 
observed to greater depth to attempt to secure detections.
The unweighted mean of their sample is 
0.58$\pm$0.36mJy).
However it is difficult, with such a small sample, to draw any firm 
conclusions, and we emphasise that investigation of the millimetric properties
of GRB hosts is ongoing, the present dataset representing merely
a pilot study.

\begin{figure}
\psfig{figure=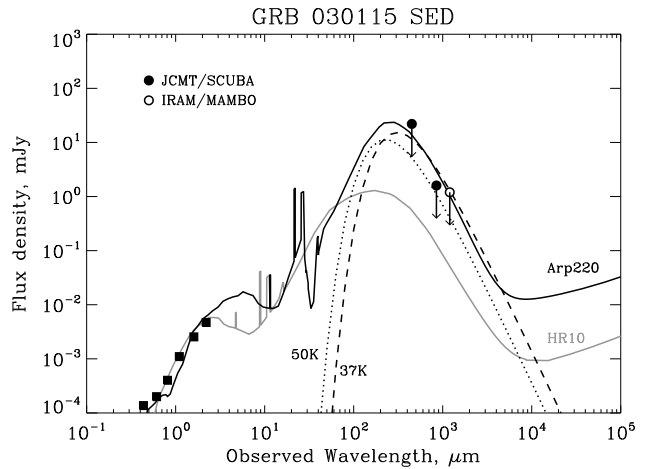,width=90mm}
\caption{Broad-band SED of the host galaxy of GRB030115, 
showing optical and near-infrared photometry (squares: Levan et al. 2006) 
together with SCUBA submm (filled circles) and MAMBO mm (unfilled circle) 
upper limits (2$\sigma$).
For comparison, model SEDs of the template galaxies,  
the prototype ultraluminous infrared galaxy Arp220 and
the prototype extremely red object HR10,
redshifted to the rest-frame of GRB030115 (assuming the photometric
redshift $z$=2.50 derived by Levan et al. (2006))
has been plotted. The models were produced by the stellar spectral evolution
code {\sc GRASIL} of Silva et al. (1998), normalised to the 
NIR/optical flux of 030115.
Our submm data indicate that GRB 030115 is marginally inconsistent with
this Arp220-like SED, although an HR10-like SED cannot be ruled out. 
Also plotted are two isothermal SEDs normalised so as to have the far infrared
luminosities implied by the optical data and the inferred extinction.
The dashed line has $T$=37K, $\beta$=1.5, canonical values commonly assumed 
for submillimetre galaxies, but is ruled out by our submm limits.
The dotted line shows that to be consistent with the data,
hotter dust is required (in this case $T$=50K).
Observations in the mid-infrared, for example with {\it Spitzer},
would be required to fully constrain any hot dust component.
}
\end{figure}

Two notable hosts we now discuss individually.

\subsubsection{GRB 030115}
Coadding all the data obtained for GRB 030115, we find
flux densities 0.0$\pm$0.8mJy at 850$\mu$m, 7$\pm$11mJy at 450$\mu$m and
0.0$\pm$0.8mJy at 1.2mm. We note that this GRB was also observed
using MAMBO by Bertoldi et al. (2003), on 20030116 and 20030118 (i.e. shortly
after the burst), to attempt to detect the afterglow. Their non-detection
(0.4$\pm$0.9mJy) can, again, be combined with our new data, to yield a
total flux 0.2$\pm$0.6mJy.
Thus although this host was not detected, we nevertheless possess reasonably 
strong upper limits at all three wavelengths. (We stress, however, that
the 450$\mu$m measurement in particular carries a substantial calibration
uncertainty. Such a deep short-wave limit is rare, but it must be
used with caution.) The 850$\mu$m RMS, in particular, would have been easily
sufficient to have detected the three submm-bright GRB hosts known to
date (e.g. Tanvir et al. 2004).

This object is of particular importance in understanding 
the host galaxies of GRBs. As noted, it is the reddest GRB host observed
to date, with a colour $R-K\approx5$ (Levan et al. 2006: hereafter L06) 
that qualifies
it as an ERO (Extremely Red Object). The afterglow, too, is exceptionally
red ($R-K\approx6$), providing further evidence for intrinsic extinction.
Although no spectroscopic redshift was measured, L06 
determine a photometric redshift $z=2.5\pm0.2$.
Adopting the calibration determined by Meurer, Heckman \& Calzetti (1999) 
for local starburst galaxies, 
the extinction implied by the rest-frame ultraviolet slope can be
used to estimate the far infrared luminosity from the observed
optical flux, assuming that the absorbed UV photons are reradiated
in the FIR/submm. The extinction at 1600\AA\ is estimated this way to be
5.3, giving a predicted $L_{\rm FIR}\approx$2.4$\times$10$^{12}$L$_{\odot}$.
Depending on the adopted stellar (IMF) initial mass function and 
star formation history, this luminosity implies a formation
rate of massive stars in the range $\approx$100--500M$_{\odot}$yr$^{-1}$.
Converting the luminosity to a predicted submm flux depends upon the
assumed dust temperature, but for a ``typical'' ULIRG SED ($T$=40K, 
$\beta$=1.5), the predicted 850$\mu$m flux is $\approx$2.5mJy. This
is inconsistent with our measurement.

In Figure 1 we plot the optical and near infrared photometry 
for this galaxy, together with the new mm and submm upper limits.
For comparison, we have also plotted (shifted to the rest frame of
GRB 030115) SEDs of the canonical
ultraluminous infrared galaxy Arp220, and the canonical extremely red
galaxy HR10.
If this Arp220 template is normalised to the optical/NIR points, it
is marginally inconsistent with the submm limits, although this represents
an extreme case, and the data cannot rule out an HR10-like SED.
The dashed line, meanwhile, shows an isothermal SED ($T$=37K, $\beta$=1.5) 
having the FIR luminosity predicted from the optical/NIR flux and inferred
extinction. Again this is inconsistent with the submm limits, but can be
accommodated if the dust temperature is increased (dotted line).

\subsubsection{GRB 020124}
Another noteworthy member of the sample is GRB 020124. 
Spectroscopy of its afterglow
revealed a high column-density ($N_H$=10$^{21.7}$) 
Dampled Lyman Alpha (DLA) system at $z$=3.2
(Hjorth et al. 2003b).
Nonetheless, the host galaxy was not detected in deep optical 
searches with {\it HST}, down to $R>29.5$ (Berger et al. 2002). 
One explanation for this faintness is that the host galaxy is
dust-rich, and therefore plausibly a mm/submm source. 
Against this interpretation, the small extinction ($A_V<0.2$)
inferred from afterglow reddening implies a low gas--dust 
(Hjorth et al. 2003b), although it is possible that dust in the vicinity
of the GRB is destroyed by its intense, beamed radiation 
(Waxman \& Draine, 2000)
hence the line of sight to the afterglow may not be representative of the
host galaxy as a whole.
Nevertheless, our 1.2mm upper limit constrains the possible 
dust mass of the host galaxy to $\la10^8$M$_{\odot}$
(varying inversely with assumed dust temperature),
which would tend to disfavour a highly extinguished host galaxy.


\section{Modelling the GRB Host Galaxy submm flux distribution}
In this section we explore the degree to which existing and future
submm observations might be able to constrain the efficacy of GRBs as
tracers of the star formation rate. 
Using models of the luminosity and redshift distributions of both 
SMGs and GRBs, we can predict the submm flux distribution that would
be obtained for GRB-selected galaxies.

\subsection{Method and assumptions}
A preliminary calculation along these lines was first performed 
by Ramirez-Ruiz, Trentham \& Blain (2002).
We adopt their assumptions concerning the properties and evolution
of SMGs, as
described by the models of Blain et al. (1999, 2001).
Briefly: the submm luminosity function is based on the local
60$\mu$m luminosity function (Saunders et al. 1990), with
luminosity evolution described by 
$\Phi(L,z)=n(z)\phi(L/g(z))$.
All submm galaxies are assumed to have an identical isothermal dust SED,
whose temperature is a parameter that can be determined by
insisting that counts in the mid-infrared are jointly fit.
When an emissivity index $\beta$=1.5 is assumed, a best fit temperature
$T$=37K results 
from fits to {\it ISO} and SCUBA counts
(See Blain et al. 1999 and Blain 2001 for full details.)

To model the properties and statistics of GRBs,
we first of all assume that the gamma-ray spectrum is described by
a Band (1993) function, using values of the spectral indices 
$\alpha$=$-$1, $\beta$=$-$2 and cut-off energy $E_0$=200keV (rest-frame).
We investigate two common parametrizations of the peak luminosity
distribution: (1) a log-normal luminosity function (specified
by a mean luminosity $L_0$ and a width $\sigma$),
and (2) a Schechter (1976) function (specified by a characteristic
luminosity $L_{\ast}$ and an index $\gamma$).
In each case,
$\rho(z)$, the (comoving) GRB rate density as a function of redshift 
is derived from the global star formation rate density,
$\rho_{\rm GRB}(z)=\eta_{\rm GRB}\times\psi_{\ast}(z)$---
e.g. as calculated from the submm models.
Similarly, the GRB rate per galaxy is assumed to be proportional to the
galaxy's far-infrared luminosity. 
Initially, we assume that $\eta_{\rm GRB}$--- the ``efficiency''
of GRB production--- is a constant.
However, in general we may consider cases where $\eta$ is a
function of redshift, or of host galaxy properties (see below).
The parameters $L_0$, $L_{\ast}$, $\sigma$ and $\gamma$ 
are then determined, for each
star formation history, by fitting the flux distribution 
of long-duration ($t_{90}>2$s) GRBs in the BATSE 4B catalogue
(Paciesas et al. 1999). 
For this purpose, we adopt a limiting photon flux sensitivity
0.27 ph cm$^{-2}$ s$^{-1}$, corresponding to the median of the BATSE 
sensitivity distribution determined by Guetta, Piran \& Waxman (2005).

\subsection{Results}

\subsubsection{Flux distribution}
In Figure 2, we show the predicted cumulative fraction of GRB hosts above 
a given flux density, for a range of submm/mm wavelengths.
In this case, we have 
assumed a detector sensitivity appropriate for {\it Swift}/BAT,
based upon the intercomparison between BAT and BATSE made by
Band (2006).
These results correspond to a log-normal GRB luminosity distribution.
However,
calculations for a Schechter function give very similar results.

\begin{figure}
\epsfig{figure=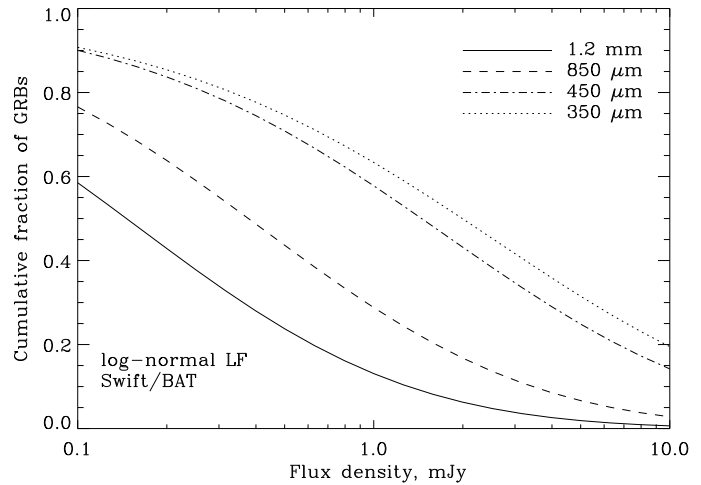,width=90mm}
\caption{Predicted fraction of GRB hosts brighter than a given flux density,
plotted for four different wavelengths in the submm/mm regime. 
This model displayed here is 
assumes a log-normal high-energy luminosity distribution for long-duration
GRBs 
(parameters estimated by fitting to the peak flux distribution of the
BATSE-4B catalogue), and
is calculated for the sensitivity of {\it Swift}/BAT. However, varying 
these parameters does not have a dramatic effect upon the results, at least
compared with the other uncertainties involved (for example, the SED assumed
to describe submm galaxies and GRB hosts).}
\end{figure}

In general,
the GRB descriptors appear not to affect the
{\it relative} numbers significantly (so long as they represent
reasonable fits to the number counts).
The adopted submm properties have a larger effect.
In particular,
we have assumed that GRB host galaxies share a common, isothermal
SED--- an SED moreover identical to that of submm galaxies. 
In reality it is possible that the mean dust temperature of GRB hosts
is different from the 37K assumed here, and that across the
sample a distribution of temperatures is to be found. 
Testing the effect of this on our predictions is not
trivial, since the dust temperature is a parameter of the Blain et al. 
(1999) submm galaxy evolution models.
A detailed refit of the SCUBA counts is beyond the scope of this 
present work, but for now we can obtain a simple indication of
the effect of varying the temperature by taking note
of the correlations between the
uncertainties in the model parameters of Blain et al. (1999), and rerunning
our calculation using the new parameters. 
Results for a plausible temperature range are shown in Figure 3.
Although the uncertainty is probably somewhat exaggerated (because it
assumes all submm galaxies are affected in the same way) the effect is much 
larger than that due to any uncertainty in the GRB properties. 
The submm SEDs of GRB hosts are ill-enough constrained that it seems plausible
that a hotter than average dust temperature--- as our limits for GRB 030115
would imply--- could account for the paucity of detections to date.

\begin{figure}
\epsfig{figure=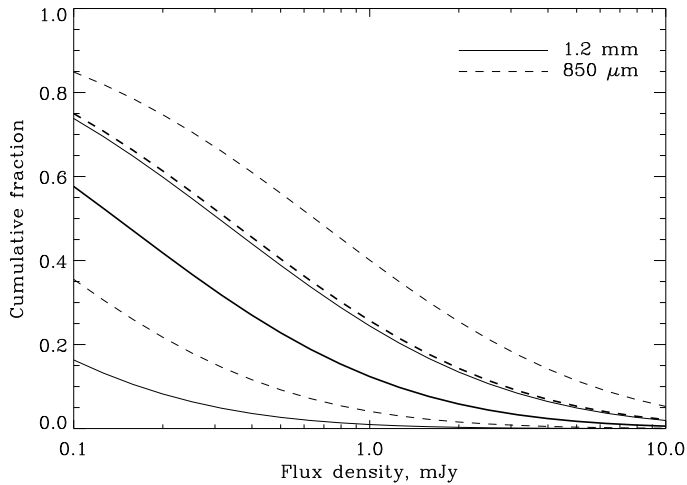,width=90mm}
\caption{Effect of varying the assumed temperature on two of the 
wavelengths plotted in Figure 2.
The bold lines correspond to the ``best fit'' temperature
as used in Figure 2, while light lines illustrate a plausible range of
temperatures (30--50K, top to bottom). 
Recall that the submm upper limits on the host
of GRB 030115 imply a temperature toward the upper part of this range. 
In reality, both GRB hosts and the
submm galaxy population in genernal are likely to exhibit a distribution
of temperatures: this calculation assumes one SED for all.
A hotter than average temperature for the GRB hosts might explain the
paucity of submm detections to date.
}
\end{figure}

At 1.2mm, we would expect to have detected $\sim$10--15 percent of a sample
with RMS $\sim$0.5--0.6mJy, assuming the $T$=37K model. This could increase
to as much as $\sim$20--25 percent if the temperature were permitted to
be as low as 30K, but would become negligible at temperatures as high as
50K. 
The predicted flux distributions enable us to calculate the 
average flux density of a large sample, to compare with the coadded
(detected plus non-detected) fluxes from observations.
The 37K model predicts $<S_{850}>\approx0.8$mJy, $<S_{1.2}>\approx0.4$mJy. 
Recall that Tanvir et al. (2004) find stacked fluxes 0.93$\pm$0.18 (weighted
mean) and 0.58$\pm$0.36 (unweighted), both consistent with this 
prediction. 
Our 1.2mm mean, on the other hand, is marginally inconsistent
with the prediction.
Our sample of five is, however, too small to confirm or reject any
of the models, but continued study of homogeneously-selected 
host samples 
should improve the constraints.
Ultimately, greater sensitivity will be attained by taking
advantage of forthcoming facilities such as ALMA (or even existing 
facilities such as {\it Spitzer}): then it will be possible to 
reach limits deep enough to discriminate between models.

\subsubsection{Predicted redshift distribution}
Figure 4 compares the predicted redshift distribution of all GRBs
(thick solid line) with 
observed, spectroscopically-derived redshifts (light-shaded histogram).
Also shown are: the variation of the expected fraction of 
submm-bright hosts with redshift, 
$\frac{d}{dz}n(S_{850}>3mJy)/\frac{d}{dz}n(total)$
(thick dashed line); 
and the redshift distribution of
the submm-observed sample (dark-shaded histogram).
Some care must be taken when interpreting the observational data, since
the observed distribution is derived from a rather inhomogeneous input
sample. Not only does the sample consist of bursts detected by a 
range of missions, but, insisting upon spectroscopic follow-up
inevitably introduces strong biases--- for example, toward lower-redshift
bursts, toward those that are intrinsically brighter,
or toward those suffering less dust extinction from their hosts.

From the figure it is clear that
most of the existing submm-observed GRB sample lies at lower redshift
than the predicted peak in the submm-bright fraction--- and indeed the
observed peak of the redshift distribution of submm galaxies (Chapman et al.
2003). This redshift bias is, therefore, another possible explanation
of the lack of submm detections in the existing host sample.
Now, however, afterglow redshift determination is more systematic, with
the accurate localisation provided by the XRT and UVOT instruments
on board {\it Swift}, and rapid ground-based follow-up via a suite of
robotic and semi-robotic telescopes. 
Submm/mm follow-up of samples resulting from such campaigns
is likely to place much more secure constraints on the star-forming
properties of GRB host galaxies than has been possible hitherto.

\begin{figure}
\epsfig{figure=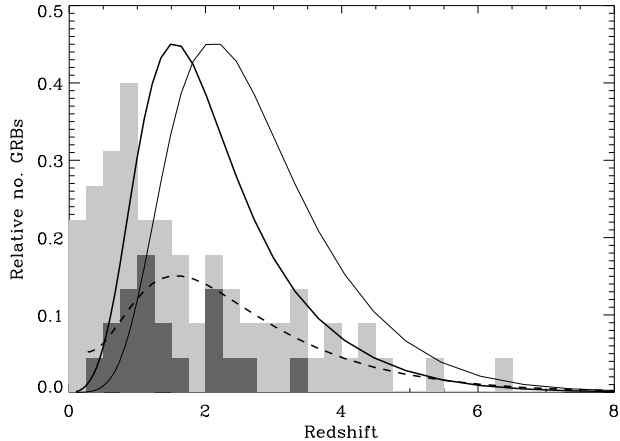,width=90mm}
\caption{Predicted (curves) and observed (histograms) 
redshift distribution of GRBs. The thick, continuous curve 
is the prediction from the basic submm galaxy model, assuming
a detector sensitivity appropriate for {\it Swift}/BAT. 
The thick dashed line illustrates the variation with redshift of 
the fraction of submm-bright ($S_{850}>3$mJy) hosts.
The light-shaded histogram shows the distribution of all GRBs with
spectroscopically-confirmed redshifts, while the dark-shaded histogram
shows the subset of this sample for which sensitive submm photometry
has been carried out. 
Finally, the light continuous curve is a variation on the basic
model incorporating a dependence of the GRB formation efficiency 
on the global average metal abundance. 
The peak is, as one would expect, shifted to higher redshift,
resulting in a distribution more consistent with that emerging
from spectroscopic follow-up of {\it Swift} GRBs.
(N.B. the vertical scale is appropriate for the dashed line (submm-bright
fraction): the remaining curves and histograms are scaled arbitrarily.) 
}
\end{figure}

\subsubsection{Dependence of GRB rate on metallicity}
One important factor that might ultimately mitigate against the
formation of GRBs in dust-rich galaxies is the role that metallicity
is thought to play (e.g. Fruchter et al. 2006). 
According to the ``collapsar'' model 
(MacFadyen \& Woosley, 1999), a low metal abundance allows the
progenitor to retain a high mass and angular momentum,
favouring the production of a black hole and accretion disk.
High detection rates of Lyman-$\alpha$ emission from GRB hosts
(Fynbo et al. 2003) could be taken as evidence that these systems
are indeed metal-poor.
If this metallicity dependence is correct, it hold consequences both
negative and positive for the use of GRBs as star formation indicators.
Whilst, on the one hand, complicating the conversion between GRB
and star formation rate, it suggests that GRBs may instead be the 
ideal means of pinpointing metal-poor galaxies--- in particular
low-mass, unenriched systems at the highest redshifts which  
are most likely to be missed in other surveys.

It is therefore important to explore the possible
effects of a metallicity dependence. To do so, we place a redshift
dependence on the 
GRB rate density---SFR density conversion factor $\eta_{\rm GRB}(z)$. 
The evolution of the average metallicity with redshift is given by the
submm galaxy models (Blain et al. 1999). 
This is converted to a relative efficiency of GRB production
using an {\it ad hoc} recipe--- which ultimately may be unrealistic, 
but, in the absence of any compelling observational or theoretical
guidelines, it serves amply to illustrate the effects. 
A sample predicted redshift distribution is plotted in Figure 4 (thin curve).
As expected, the peak is shifted toward higher $z$ where the 
average abundance of heavy elements is smaller.  
However, without separately encoding galaxy-to-galaxy variations in 
the metal abundance, the effects on the submm flux distribution are small.

This calculation is, we emphasise, only illustrative at present.
For example, it may be more appropriate, for GRB hosts, to consider
metallicities traced by optical galaxy surveys, rather than 
submm surveys as used here. 
As the redshift distribution of GRBs becomes more fully sampled,
(for example via spectroscopic follow-up of large samples of
{\it Swift} bursts), it will soon be possible to place constraints on
a wider range of models in this way.

\section{Summary}
Following from the previous study of Tanvir et al. (2004),
we have further investigated the millimetre/submillimetre properties
of the host galaxies of GRBs, in order to characterise the efficacy
of GRBs as star formation indicators. 
Specific increments over the T04 study include:
(1) we have conducted the first survey of GRB hosts at millimetric
wavelengths, 
with the MAMBO2 bolometer array on
the IRAM 30m Pico Veleta telescope. 
None of these targets was detected, down to an average RMS
$\approx$0.6mJy at 1.2mm;
(2) we obtained deep submm photometry of
GRB030115, whose high intrinsic extinction inferred from 
its optical/NIR spectral slope make it a promising candidate 
submm galaxy. Despite its ERO-like optical colours, however, 
this galaxy is not 
detected in the mm/submm, to deep limits at 
850$\mu$m ($\sigma$=0.8mJy) and 450$\mu$m ($\sigma$=11mJy);
(3) we have modelled the redshift and flux distribution of GRB hosts,
assuming a link between GRBs and the submm galaxy population.
A novelty of these models is that they  
take account of the metallicity bias widely proposed to 
affect the GRB-to-star formation rate conversion.
As such they potentially have much wider applicability than 
the derivation of submm properties, and we will further develop
these ideas in future publications (Priddey et al., in prep.).

The non-detection of GRB 030115 is revealing.
One might contrast this result with the three GRBs that {\it do} possess
submm detections, for their optical/NIR colours are much bluer.
The broadband spectrum of the GRB 030115 host is inconsistent with 
the SED of an extremely luminous infrared galaxy such as
Arp220 or with a cool, isothermal model, but 
hotter dust ($\ga$50K), or template SEDs of other submm-luminous 
galaxies, 
cannot be ruled out.
Observation in the mid-infrared with missions such as {\it Spitzer}
should also be able to constrain any hot dust component
too faint to be seen in the submm.

We emphasise that this work is ongoing: 
in the imminent future we will be able to draw upon larger,
post-{\it Swift}
samples of GRBs to ensure a uniform sample selection--- enabling,
for example, a more uniform redshift distribution.
For the moment, it seems that the trend of a low submm detection rate
of GRB hosts, seen in previous surveys, is maintained. 

What are the implications of a low mm/submm detection of GRB hosts?
We have shown that there is sufficient uncertainty in models and
underlying assumptions, as yet poorly constrained by observation 
(for example the adopted dust temperature) that a correlation between
massive, dust-enshrouded star formation and GRB production cannot be
firmly ruled out. 
Sample selection biases (e.g. against high redshift and highly
extinguished bursts) are also likely to have played a significant
role in previous studies. Our models indicate that redshift bias
in particular could account for the lack of detections within
existing surveys.
The new observations reported here (5 hosts all at $z>1$, one highly
extinguished host and one extremely faint host) were taken in part to 
alleviate such problems. 
Prior to ALMA,
observations of consistently followed-up samples with existing 
facilities (e.g. IRAM-30m, APEX) must be made to enable us to
make further progress in exploring these effects. 
The capabilities of {\it Swift}, combined with efficient ground-based
follow up, show promise in being able to yield such a sample.

\section{Acknowledgments}
RP thanks PPARC for support. We are grateful to those observers and support
staff at JCMT and IRAM Pico Veleta who carried out observations for us in
service mode, and we thank the anonymous referee for constructive comments. 
The JCMT is operated by the Joint Astronomy Centre in Hilo,
Hawai`i, on behalf of the parent organisations: the Particle Physics and
Astronomy Research Council in the United Kingdom, the National Research
Council of Canada and the Netherlands Organisation for Scientific Research.

\appendix

\label{lastpage}

\end{document}